# MBCT: A Monero-Based Covert Transmission Approach with On-chain Dynamic Session Key Negotiation

Zhenshuai Yue, Haoran Zhu, Xiaolin Chang, *Senior Member*, *IEEE*, Jelena Mišić, *Fellow*, *IEEE*, Vojislav B. Mišić, *Senior Member*, *IEEE*, Junchao Fan

*Abstract*—Traditional covert transmission (CT) approaches have been hindering CT application while blockchain technology offers new avenue. Current blockchain-based CT approaches require off-chain negotiation of critical information and often overlook the dynamic session keys updating, which increases the risk of message and key leakage. Additionally, in some approaches the covert transactions exhibit obvious characteristics that can be easily detected by third-parties. Moreover, most approaches do not address the issue of decreased reliability of message transmission in blockchain attack scenarios. Bitcoin- and Ethereum-based approaches also have the issue of transaction linkability, which can be tackled by Monero-based approaches because of the privacy protection mechanisms in Monero. However, Monero-based CT has the problem of sender repudiation.

In this paper, we propose a novel *M*onero-*B*ased *CT* approach (MBCT), which enables on-chain session key dynamically updating without off-chain negotiation. MBCT can assure non-repudiation of transmission participants, confidentiality of keys, reliability of message transmission and less observable characteristics. There are achieved by the three components in MBCT, namely, a sender authentication method, a dynamically on-chain session key updating method and a state feedback method. We implement MBCT in Monero-0.18.1.0 and the experiment results demonstrate its high embedding capacity of MBCT.

## I. INTRODUCTION

COVERT transmission (CT) aims to provide a more concealed channel to transmit data between sender and receiver. CT data is usually hided in normal data flows in order to avoid being perceived by third parties. It is widely used in various applications like satellite communication [1], wireless communication [2], and Internet of Things [3]. There are at least the following three issues with traditional CT approaches: 1) Centralized facility dependence. The extensive utilization of TCP/IP protocol stack makes traditional approaches susceptible to potential malfunctions in central infrastructures. 2) Channel detection. Traditional approaches are prone to detection [4], which makes covert channels vulnerable to exposure and attacks. 3) Identity exposure. Due to the inherent structure of the communication protocol, the network information of both parties may be leaked to attackers.

To solve the above issues, blockchain offers excellent channel support for CT thanks to its decentralization, flood broadcasting and identity anonymity features [5]. In particular, the distributed architecture ensures that each full node retains a full copy of whole data, which prevents the blockchain from performance degradation due to single node failures. The flood broadcast improves the concealment of CT. The anonymous user identity in the blockchain avoids leakage of real network addresses. Moreover, the large number of transactions circulating in the blockchain network provides an excellent information carrier for CT. A blockchain-based CT has the following six major steps, shown in Fig. 1. ① The sender encodes covert messages. ② The sender embeds encoded messages into a transaction $TX_c$. ③ The sender connects to blockchain network and broadcasts $TX_c$ to the blockchain network until $TX_c$ is shown on chain. The covert transactions and normal transactions should be indistinguishable to third parties. ④ The receiver filters covert transaction $TX_c$ by traversing recent blocks. ⑤ The receiver extracts embedded covert message. ⑥ The receiver decodes covert messages.

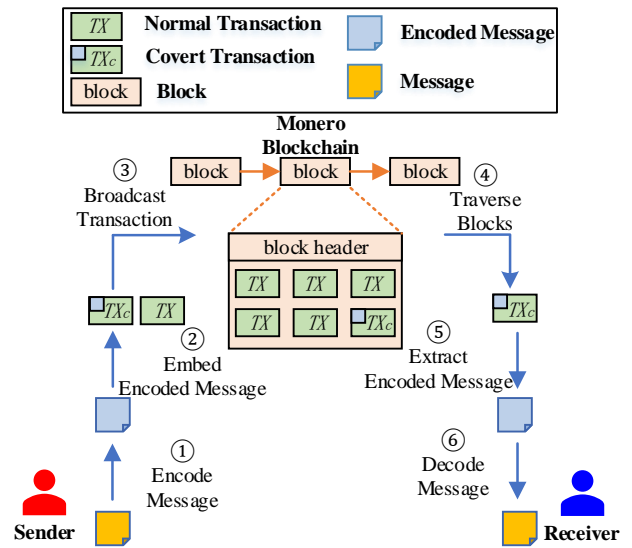

Fig. 1. Blockchain-based covert transmission

There are numerous CT approaches based on various public blockchains. But the CT approaches only for Bitcoin [6]-[13] or only for Ethereum [14]-[17] or for both Bitcoin and Ethereum [21]-[26] cannot achieve fully anonymous. It is due to that multiple transaction addresses can be associated with a same entity [28][29], leading to identity exposure as in traditional CT approaches. Monero [30] is another well-known public blockchain that shields users' identities through advanced privacy protection mechanisms like ring signature, stealth address and confidential transaction. The real address

• Zhenshuai Yue, Haoran Zhu, Xiaolin Chang and Junchao Fan are with the School of Computer and Information Technology, Beijing Jiaotong University, Beijing 100044, China (e-mail: {22120512,21112051, xlchang,23111144}@bjtu.edu.cn).
• Jelena Mišić and Vojislav B. Mišić are with the Department of Computer Science, Toronto Metropolitan University, Toronto, ON M5B 2K3, Canada (e-mail: {jmisic,vmisic}@torontomu.ca).

and amount of a Monero transaction are concealed. These mechanisms solve the issue of identity exposure. Existing Monero-based CT approaches embed messages into the irrelevant public keys of ring signatures [18][20] and transaction amounts [19]. Due to Monero's anonymity, the receiver cannot verify the identity of the sender, and the sender can also deny the communication behaviors. In summary, the existing blockchain-based CT approaches cannot defend against the following five threats detailed in Section II.B.
  i) Information leakage of off-chain negotiation.
  ii) On-chain session key leakage.
  iii) Malicious sender denial of transmission behavior.
  iv) Integrity compromised by abnormal state.
  v) Channel exposure due to transaction characteristics and associations

The above discussions motivate our work. In this paper, we explore to tackle the above issues by proposing a *Monero-Based CT* approach (MBCT). MBCT can assure non-repudiation of transmission participants, confidentiality of keys, reliability of message transmission and less observable characteristics. There are achieved by the three components in MBCT, namely, a sender authentication method, a dynamically on-chain session key updating method and a state feedback method. To the best of our knowledge, we are the first to propose a CT mechanism implementing on-chain dynamic session key updates without off-chain negotiation.

The innovations of our approach are listed as follows.

1) We propose a sender authentication method, which can maintain the anonymity of sender while ensuring the non-repudiation of sender. By embedding the sender's digital signature into the stealth addresses of Monero transactions, the receiver can authenticate the sender's identity. During subsequent transmission process, the digital signature is used in the process of generating masked amount, ensuring the sender's non-repudiation. Since the embedded digital signature is imperceptible and inaccessible to third parties, the sender's anonymity is then maintained.

2) We propose a dynamically on-chain session key updating method, which not only guarantees independency of each session but also requires no off-chain negotiation, isolating off-chain risks. The on-chain session key changes based on the stealth address variations in each transaction, ensuring that sessions are independent of each other and preventing a single session key leak from affecting other sessions.

3) We propose a state feedback method, in which the sender can perceive the communication status covertly and resend lost messages timely due to reasons such as blockchain attacks, maintaining the integrity of the communication.

We conduct a comprehensive analysis of our approach's security, concealment and embedding rate, and experimentally test the embedding and extraction efficiency and validate its stealthiness. The analysis and experiment results demonstrate that our approach is secure and stealthy, featuring high embedding rate.

The rest of this paper is organized as follows. We present the background and related work in Section II. System overview and a detailed description of MBCT is presented in Section III. Security analysis and performance evaluation are described in Section IV. Section V provides the conclusion.

## II. BACKGROUND AND RELATED WORK

This section first presents the background and then introduces the related work about blockchain-based CT.

### A. Background

*1) Diffie-Hellman key exchange with elliptic curves*

An elementary shared secret exchange using the Elliptic Curve Diffie-Hellman (ECDH) exchange method between two parties, Alice and Bob, could proceed as follows.

Alice and Bob generate their own private-public key pairs $(k_A, K_A)$, $(k_B, K_B)$. In this process, the public keys $K_A$ and $K_B$ are openly shared, while each party securely maintains their own private key (i.e., $k_A$ and $k_B$). To elaborate further, the private key $k$ is a random number selected within the range $0 < k < l$, where $l$ is a sufficiently large number. The public key $K$ is a coordinate point on the elliptic curve, calculated using the following equation.

$$K = kG, \quad (1)$$

where $G$ is the base point of the elliptic curve. This setup ensures that the public key can be easily derived from the private key. However, due to the elliptic curve discrete logarithm problem (ECDLP), it is computationally infeasible to reverse the process and determine the private key from the public key.

Equation (2) gives the detail of calculation for ECDH:

$$S = k_A K_B = k_A k_B G = k_B k_A G = k_B K_A, \quad (2)$$

where $S$ is a coordinate point on the elliptic curve as a shared secret. This shared secret can be used to generate stealth address of Monero transaction. The security of the exchange is based on the ECDLP, which makes determining the shared secret from the public keys alone computationally prohibitive.

*2) Stealth address of Monero transaction*

Monero blockchain [27] is designed to protect user privacy. As one of the privacy protection mechanisms of Monero, stealth address mechanism allows each Monero transaction to generate a new public address for receiving funds. This means that even though outside observers can see transaction records on the public blockchain, they cannot associate the receiving address in the transaction with any specific user or previous transaction.

To generate a stealth address for receiver, the sender first selects a 256-bit random number $k^r$ as transaction private key, and then uses the receiver's public view key $K_B^v$ to generate a shared secret $S$, as illustrated in Eq. (3):

$$S = k^r K_B^v, \quad (3)$$

and then the sender uses the receiver's public spend key $K_B^s$ to generate stealth address, as illustrated in Eq. (4):

$$K = \mathcal{H}(S,t)G + K_B^s, \quad (4)$$

where $K$ is the stealth address, $\mathcal{H}$ is a cryptographic hash function and $t$ is the non-change output index stored in the transaction in plaintext.

In the same way, the receiver calculates the shared secret $S$ due to ECDH key exchange as illustrated in Eq. (5):

$$S = k^r K_B^v = k^r k_B^v G = k_B^v k^r G = k_B^v K^r, \quad (5)$$

where $k_B^v$ is the private view key of the receiver and $K^r$ is the public transaction key stored in the transaction in plaintext. Then the receiver could generate the same stealth address as illustrated in (4). Therefore, the sender and receiver reach a consensus on the same stealth address.

*3) Masked amount of Monero transaction*

In Monero, the confidential transaction ensures the legality of transactions while using masked amounts to hide the actual amounts based on Pedersen Commitments and Range Proofs, thereby protecting transaction privacy. We focus on masked amounts to align closely with our proposed approach, rather than on Pedersen Commitments and Range Proofs.

The sender first calculates a shared secret $S$ as shown in Eq.(3). Then the masked amount of real amount $a$ for non-change output index $t$ is generated by Eq. (6):

$$h = a \oplus_8 \mathcal{H}("amount", \mathcal{H}(S,t)), \quad (6)$$

where $h$ is the masked amount, "amount" is a string and $\oplus_8$ means to perform an XOR operation between the first eight bytes of each operand. Subsequently, after receiving the transaction, the receiver first calculate the same shared secret $S$ through (5) and then calculate the real amount by Eq. (7):

$$a = h \oplus_8 \mathcal{H}("amount", \mathcal{H}(S,t)). \quad (7)$$

Our approach utilizes stealth addresses to transmit information and leverages hiding amounts for anti-forgery and conveying message sequences.

*4) Blockchain-Based CT*

In blockchain, each full node maintains a copy of the block data and achieves distributed consensus like Proof of Work (PoW). This mechanism ensures that modifying data on the blockchain requires more than 51% of the total network hashing power, which grants the blockchain a high degree of reliability of message transmission.

Blockchain-based CT refers to covert transmission between parties utilizing the characteristics of blockchain technology. There are commonly two kinds of blockchain-based covert channels, including blockchain-based covert storage channels (BCSC) and blockchain-based covert timing channels (BCTC). BCTC utilizes timing sequence characteristics to embed information, making it susceptible to network quality interference, thus exhibiting poor reliability of message transmission. BCSC employs storage fields on the blockchain to embed covert information. The decentralized nature of blockchain ensures that this channel has higher reliability of message transmission. Our approach utilizes Monero BCSC channel.

*B. Threat Model and Security Requirements*

In current blockchain-based CT approaches, there are typically two phases: negotiation and transmission. In the negotiation phase, the communicating parties usually use an off-chain channel to negotiate keys, filtering labels and other necessary parameters. In the transmission phase, the sender processes the message using the pre-negotiated keys and coding rules, and transmit it through carrier on blockchain; the receiver filters and extracts the information using pre-negotiated filtering labels and keys. There are five types of threats (THs) concerning negotiation and transmission:

1) **TH1: Information leakage of off-chain negotiation.** Before on-chain transmission, attackers might target the off-chain channel to obtain critical negotiation parameters through network attacks such as sniffing, phishing, and man-in-middle attacks. If negotiation information is leaked, subsequent on-chain covert communication cannot proceed as usual.
2) **TH2: On-chain session key leakage.** To ensure message confidentiality, messages are encrypted prior to transmission. If the encryption key remains unchanged for an extended period and is compromised, it could lead to the exposure of the associated session information.
3) **TH3: Malicious sender denial of transmission behavior.** In Monero-based CT, due to Monero's ring signature mechanism, the receiver cannot determine the sender's real address from the transaction data, posing a risk of the sender denying the transmission activity.
4) **TH4: Integrity compromised by abnormal state.** Although blockchain has a high reliability of message transmission, attacks on the blockchain may prevent normal transactions from being propagated. If the sender is unaware of the transaction loss and the receiver does not know whether the information has been received completely, the integrity of the communication is compromised.
5) **TH5: Channel exposure due to transaction characteristics and associations.** The receiver filters covert transactions from numerous transactions using labels. As the number of transactions increases, an adversary may expose the covert channel by statistically analyzing transaction characteristics and associations.

To defend against these threats, MBCT at least needs to meet the following security requirements (SRs). TABLE I lists the correspondence between threats and security requirements.

TABLE I

CORRESPONDENCE BETWEEN THREATS AND SECURITY REQUIREMENTS

|     | TH1 | TH2 | TH3 | TH4 | TH5 |
| --- | --- | --- | --- | --- | --- |
| SR1 | √   | √   |     |     |     |
| SR2 |     |     | √   |     |     |
| SR3 |     |     |     | √   |     |
| SR4 |     |     |     |     | √   |
| SR5 |     |     |     |     | √   |

1) **SR1: Confidentiality of on-chain session key.** To ensure the confidentiality of CT, it is crucial to maintain the confidentiality of session keys, which involves two main aspects. The first aspect is to avoid off-chain negotiations and then effectively reduce the impact of off-chain attacks. The second aspect is to dynamically update session keys on-chain, which can prevent the leakage of all related sessions due to long-term key exposure.
2) **SR2: Non-repudiation of transmission parties.** Transactions are stored on chain in plaintext during CT based on Bitcoin and Ethereum. However, during the process of CT based on Monero, additional measures are required to ensure the transmission parties are accountable, preventing them from denying the communication behavior.
3) **SR3: Reliability of message transmission under blockchain attack.** CT needs to ensure reliable message transmission to maintain its availability, which is mainly reflected in two aspects. Firstly, the immutability of the blockchain provides inherent reliability. Secondly, the transmission approach needs to offer additional reliability in the event of an attack on the blockchain.
4) **SR4: Unlinkability of covert transactions.** To prevent an adversary from associating other covert transactions through one leaked covert transaction, it is necessary to ensure unlinkability between covert transactions. This includes both direct links, where an adversary can directly connect one covert transaction to another, and indirect links, where connections are made through normal transactions associated with a covert one.
5) **SR5: Obscurity of covert transactions.** On-chain transmission needs to be stealthy, ensuring that third parties cannot distinguish between covert transactions and normal transactions, thereby reducing the risk of adversaries taking further aggressive actions.

*C. Related Work*

We have investigated the approaches for CT based on blockchain, and categorize them according to the type of blockchain, including Bitcoin, Ethereum, Monero, and those applicable to both Bitcoin and Ethereum.

*1) Bitcoin-based CT*

Bitcoin is the earliest public blockchain and is also the first to be used in research on blockchain-based CT. Partala et al. [6] first proposed a bitcoin-based covert communication approach BLOCCE and systematically proved its correctness and security. The sender embeds covert information into the last bit (LSB) of bitcoin address and the receiver filters covert transaction by pre-negotiated input address. Although the low embedding rate of this approach results in low engineering application value, it creates a precedent for subsequent research on blockchain-based CT. Tian et al. [7] proposed a CT method DLchain with dynamic label. The sender and receiver pre-negotiate encryption keys to encrypt message before inserting it. The receiver uses a random factor twice to extract message, leading to privacy leaks and arouse suspicion from third parties. Wang et al. [8] proposed a CT method that matches message sequences with address characters and records important parameters in a file sent to the recipient. This method requires prior negotiation of DES keys, and the security and concealment of the file transmission are not guaranteed. Wang et al. [9] combined transaction addresses to generate labels in the form of a label tree, avoiding the use of static labels. Despite this, they need to negotiate in advance to generate the ancestor secret key for the label tree, and if this key is leaked, it would lead to the exposure of the communication. Luo et al. [10] used the bitcoin address interaction matrix and transaction amount to embed hidden information, and used a negotiated address sequence to filter covert transactions. But this method may involve address reuse issues in transaction creation. Zhang et al. [11] proposed a derivate matrix-based covert communication method, in which the sender and receiver need to pre-share mapping rules and all private keys. Both the above methods require using a large number addresses for transactions due to low embedding rate when transmitting messages, leading to address reuse, which is vulnerable to statistical analysis. Based on hash chains and Elliptic Curve Digital Signature Algorithm (ECDSA) chains, Cao et al. [12] proposed a CT method suitable for Bitcoin. The communicating parties need to negotiate the key off-chain, then use this key to generate public keys for transactions. Unlike the methods mentioned above that use transaction creation for transmitting information, Zhu et al. [13] achieved CT by encoding the transaction hashes sequence in the *inv* and *getdata* messages between Bitcoin nodes. The premise of this method is that both parties need to negotiate information such as IP addresses, encoding rules, and embedding locations.

*2) Ethereum-based CT*

Ethereum features a unique smart contract mechanism that allows for storing and executing code on the blockchain, offering a large operational space, and serving as one of the carriers for implementing CT. A. I. Basuki et al. [14] utilized image steganography to embed information into images, then transmit the image URL and other relevant data through Ethereum. This method requires filtering based on a pre-negotiated destination address and periodically replacing the steganographic image with a normal image to protect against malicious attacks. However, this approach relies on external carriers and imposes strict restrictions on the access time for recipients. Liu et al. [15] proposed a CT method based on the Ethereum Whisper protocol, pre-negotiating AES encryption keys and mapping binary bits to transaction amounts within the Hash-based Message Authentication Code (HMAC) value range of the key, which had a low embedding rate. In addition, this method does not describe an effective filtering method, and the irrelevant transactions would potentially impact the information extraction process.

Similarly, Zhang et al. [16] have also proposed a CT method based on the Whisper protocol. Before communication, they negotiate *topic-key* pairs according to the protocol, and then use the payload in the letter body for information transmission. However, this method differs from using transactions for transmission, as it requires the sender to spend additional time and resources to compute the PoW for the message.

Based on smart contracts, Zhang et al. [17] proposed a CT method, mapping the voting options of a voting contract with the bid amounts of a bidding contract to ciphertexts. This method needs negotiate encryption keys and voting addresses with the risk of information leakage. This allows only limited information to be transmitted during a single interaction with the smart contract, requiring frequent interactions.

*3) Monero-based CT*

Monero, an emerging public blockchain in recent years, employs various privacy protection mechanisms to enhance the anonymity of transactions, meeting the concealment requirements of CT. Guo et al. [18] proposed a Monero-based transmission method, encrypting messages by pre-negotiated keys and embedding the ciphertext into the last bit of ring signature unrelated public keys. Although theoretically the embedding rate increases with the number of transaction inputs, large number of inputs of one transaction are uncommon in the network. Liu et al. [19] proposed a Monero-based CT method that considered the impact of blockchain attacks on the reliability of message transmission, reducing the impact of eclipse attacks and crawler attacks on communication. But since the information is directly encoded into the transaction amounts based on pre-negotiated message length, the embedding rate is not high. As the message length in a single transaction increases, the associated costs also rise significantly. Su et al. [20] stored files in the interplanetary file system (IPFS) and then transmitted the hashes of files via the Monero blockchain. The above three Monero-based CT approaches, due to the obfuscation of the sender's public key by ring signatures, cannot prevent the issue of the sender denying the communication action.

*4) CT applicable to both Bitcoin and Ethereum*

There are also some CT methods applicable to both Bitcoin and Ethereum. Gao et al. [21] proposed a method for CT using Kleptography, which offers high levels of concealment. However, this method needs pre-negotiate keys and requires two identical input addresses to retrieve the covert messages. With long-term use, this feature could be easily identified through statistical analysis. Zhang et al. [22] proposed a CT method with dynamic label, embedding message segments into storage fields of transactions. The communication participants need negotiate encryption keys, label keys and obfuscation parameters in advance and the receiver filters transactions by customized data field, such as OP_RETURN of Bitcoin and Input of Ethereum. But most normal transactions do not add extra data to the custom field, which could raise suspicions among adversaries. Liu et al. [23] also proposed a communication method with dynamic labels, where the message authentication code generated from pre-negotiated block height and keys serves as the private key to generate the recipient's address. This method allows only two bits of information to be embedded per address, resulting in a low embedding rate. Zhang et al. [24] proposed a group covert communication method, which achieves group transmission by dividing interaction addresses through pre-negotiated keys and an address pool. Chen et al. [25] proposed an unobservable blockchain-based covert channel, encoding messages as private keys, and filtering covert transactions by their signature and amounts, and conducted a detailed analysis of its concealment. However, the approach still requires the pre-share keys off-chain and carries a risk of information leakage. In response to the challenge of securely transmitting a master key, Zhang et al. [26] introduced a novel method for CT. This method embeds covert information into STC mappings while using secret sharing to transmit the master key, without discussing how the receiver can filter covert transactions from the numerous transactions on the blockchain. The embedding rate is low for the above two methods and needs to be further improved.

Based on the threat model described in Section II.B, all the above works [6]-[26] need off-chain negotiation and ignore the potential leakage of static on-chain session keys. Some works [18]-[20] based on Monero cannot prevent malicious senders from denying transmission actions. Most studies [6]-[18][20]-[26] have not explored mechanisms to maintain the reliability of message transmission under blockchain attacks. All work based on Bitcoin [6]-[13], Ethereum [14]-[17] and those applicable to both Bitcoin and Ethereum approaches [21]-[26] could be exposed due to statistical analysis of transaction associations, and some of these transactions [6]-[8][10][14][17][21][22][24] display distinct characteristics, further increasing the risk of exposure. To mitigate the threats to the security of CTs, we have implemented a novel CT approach with dynamic key updates, participant non-repudiation, and feedback on the status of communication, without off-chain negotiations and observable characteristics. TABLE II summarizes these works in terms of security requirements defined in Section II.B.

III. SYSTEM OVERVIEW AND MBCT APPROACH

This section first gives an overview of system and then describes the details of MBCT approach. The symbol definition is summarized in TABLE III.

A. System Overview

The system comprises three key entities, namely the sender, receiver and Monero blockchain.

- **Sender**. The sender connects to the blockchain network by running a full node or a simplified payment verification (SPV) node. During the process of building transactions, the sender embeds message into transactions and then broadcasts them to the blockchain network.

- **Receiver**. The receiver can run a blockchain node or call the service of the online node through Remote Procedure Call (RPC). He filters covert transaction by traversing recent blocks. After obtaining covert transaction, he will extract message embedded by the sender from the transaction fields.

- **Monero Blockchain**. Each time a sender creates and broadcasts a covert transaction, the miner nodes in the blockchain will package it into a new block after verifying the validity of the transaction. Eventually the transaction will be permanently stored on the chain. The sender and receiver transmit message via Monero Blockchain due to its advanced cryptographic techniques. The distributed nature of blockchain makes it with a high degree of reliability of message transmission, but when it is attacked, the transactions may get lost and reliability may decrease.

TABLE II

COMPARISON OF EXISTING BLOCKCHAIN-BASED CT APPROACH

| Ref. | Blockchain | | | Filter Field | Security Requirements | | | | |
|---|---|---|---|---|---|---|---|---|---|
| | Bitcoin | Ethereum | Monero | | SR1 | SR2 | SR3 | SR4 | SR5 |
| [6] 2018 | √ | | | Input address | | √ | | | |
| [7] 2019 | √ | | | OP_RETURN | | √ | | | |
| [8] 2020 | √ | | | Parameter file | | √ | | | |
| [9] 2023 | √ | | | Output address | | √ | | | √ |
| [10] 2022 | √ | | | Address pool | | √ | | | |
| [11] 2023 | √ | | | Address matrix | | √ | | | √ |
| [12] 2022 | √ | | | Output address | | √ | | | √ |
| [13] 2023 | √ | | | Control information | | √ | | √ | √ |
| [14] 2019 | | √ | | Destination address | | √ | | | |
| [15] 2020 | | √ | | - | | √ | | | |
| [16] 2021 | | √ | | Topic of letter | | √ | | | √ |
| [17] 2022 | | √ | | Voting address | | √ | | | |
| [18]2021 | | | √ | | | | | √ | √ |
| [19] 2022 | | | √ | One-time address | | | √ | √ | √ |
| [20] 2023 | | | √ | | | | | √ | √ |
| [21] 2020 | √ | √ | | Input address | | √ | | | |
| [22] 2023 | √ | √ | | Customized data field | | √ | | | |
| [23] 2023 | √ | √ | | Output address | | √ | | | √ |
| [24] 2022 | √ | √ | | Address pool | | √ | | | |
| [25] 2024 | √ | √ | | Signature and amounts | | √ | | | √ |
| [26] 2024 | √ | √ | | - | | | | | |
| **Ours** | | | √ | One-time address | √ | √ | √ | √ | √ |

TABLE III
NOTATIONS

| Symbol | Definition |
|---|---|
| $K_A^v$, $k_A^v$ | Monero view public-private key pair of sender |
| $K_A^s$, $k_A^s$ | Monero spend public-private key pair of sender |
| $K_B^v$, $k_B^v$ | Monero view public-private key pair of receiver |
| $K_B^s$, $k_B^s$ | Monero spend public-private key pair of receiver |
| $K^r$, $k^r$ | Monero transaction public-private key pair |
| $K_i^{ori}$ | Original stealth address of $i$-th non-change output |
| $K_i^{new}$ | New stealth address of $i$-th non-change output |
| $G$ | Base point of the elliptic curve in Monero |
| $\mathcal{H}$ | Cryptographic hash function |
| $r, s$ | EdDSA signature |
| $m$ | Message |
| $c$ | Ciphertext of message |
| $a$ | Real transaction amount |
| $h$ | Masked transaction amount |
| $t$ | Non-change output index of Monero transaction |
| $mss$ | Missing message sequence |
| $sfa, sft, sff$ | Special fields of $AuthTx$, $TransTx$ and $FbTx$ |
| $Sign$ | EdDSA signature generation function |
| $Enc$ | AES encryption function |
| $GenAmount$ | Amount generation function |
| $\|$ | Operator to concatenate two strings |
| $\oplus_8$ | Operator to perform an XOR operation between the first eight bytes of each operand |

## B. MBCT Approach Overview

We utilize **Alice** and **Bob** as the sender and receiver respectively to presents an overview of our approach. Alice and Bob use Monero blockchain to achieve the entire communication process.

The whole covert communication is split into three stages, authentication stage, transmission stage and feedback stage. For each stage, we present an embedding method to reach the goal of this stage. Since the extraction algorithm is the inverse process of embedding, we only make a comprehensive description of embedding algorithms as illustrated in Section III. C.

A normal assumption is made that Alice and Bob both possess each other's public keys, $K_A^v$, $K_A^s$, $K_B^v$, $K_B^v$ since these keys are published on the Monero blockchain.

(1) **Authentication Stage:** This stage is designed to initial the entire transmission process, serving for Bob to verify the identity of Alice. Alice firstly generates a digital signature using her secrete view key and embeds it into the stealth address of current transaction. This is referred to as the authentication transaction ($AuthTx$). Then, Bob identifies the $AuthTx$ by traversing recent transactions, extracts the digital signature and validates it for the subsequent transmission.

(2) **Transmission Stage:** This stage is designed for the secure and orderly transmission of messages. Firstly, Alice splits messages into several parts to fit the length of the transaction carriers. For each segments, a one-time session key is generated from a stealth address of each transaction, encrypting message to incorporate it into the stealth address field. Then, she generates corresponding transaction amounts based on the embedded message sequence, using the specified amounts encoding method.

We refer to the transactions mentioned above as transmission transactions ($TransTx$s). Since only Alice and Bob can calculate the stealth address, only they can identify the $TransTx$ and extract message along with its sequence correctly.

(3) **Feedback Stage:** This stage is designed to provide feedback on communication status and prevent information loss. After receiving all messages or in case of missing a message segment in a single communication, Bob embeds a digital signature generated from his secrete view key to the stealth addresses and embeds the missing segment sequence to the amount field of the Monero transaction intended to be sent to Alice. We call this kind of transaction as feedback transaction ($FbTx$). Then Alice identifies the $FbTx$ and decides whether to resend messages based on whether any messages are missing.

## C. Method Design

### 1) Sender Authentication Method

This method is implemented during the authentication stage, embedding the sender's digital signature into the stealth addresses of one Monero transaction. The digital signature is generated by the sender using the private view key as the private key and the hash value of the public transaction key of the current transaction as the message. The digital signature algorithm employs Edwards-curve Digital Signature Algorithm (EdDSA), which produces a 64-byte signature, double the size of the stealth address.

Firstly, Alice chooses a 256-bit random factor $k^r$ as private transaction key, calculates and stores $K^r$ as illustrated in (1). Then, Alice calculates hash of $K^r$ through cryptographic hash function and generates digital signature $(r, s)$ of 64 bytes. Subsequently, Alice generates two stealth addresses of 32 bytes with Bob's public keys as shown in (3) and (4), and then performs an XOR operation with each half of the digital signature separately to generate two new stealth addresses. Finally, Alice verifies whether these two new addresses are valid. If so, Alice will create an $AuthTx$ with these two new stealth addresses. Otherwise, the above process will be repeated until the stealth address is valid. We denote the stealth addresses with embedded signature as a **special field of $AuthTx$**. The generation process of the special field for $AuthTx$ is described in Algorithm 1. All stealth addresses in the algorithm are non-change output addresses, and output addresses of change are not involved in this method.

### 2) Dynamically On-chain Session Key Updating Method

In transmission stage, to avoid information leakage through off-chain negotiation, the sender uses this method to dynamically update session keys on-chain. This method leverages a Diffie-Hellman key exchange within the stealth address. It uses the stealth address and the receiver's public key to generate a hash value. Since the hash can be considered randomly distributed, we use this hash as the encryption key for the message. The encrypted ciphertext is then XORed with the original one-time address. This ensures that even if an attacker manages to associate the $TransTx$ with the public keys of the communicating parties, he cannot extract any plaintext information from the transaction.

When the message length exceeds 32 bytes, the length of a stealth address, it is necessary to segment the message. To ensure that the receiver can confirm the complete receipt of the message and accurately reassemble the message sequence, we design an amount encoding method to embed message sequences as shown in Fig. 3. To avoid generating transactions with large amounts, we use the last ten decimal places of the Monero amount, measured in atomic units called piconero, for encoding. The first digit serves as a flag; a flag of one signifies that the message segment is not the final one, whereas a flag of zero indicates the final segment. The middle six digits are populated with random numbers, and the final three digits represent the message sequence, filled from the end to the beginning.

Alice first divides message $m$ into $n$ segments of 32 bytes each, padding the final segment if less than 32 bytes with random characters. Based on amount encoding method of our method, which is shown in Fig. 3, Alice generates transaction amounts corresponding to the message sequence. Monero utilizes cryptographic hash function to generate masked amount as shown in (6). In addition, our protocol incorporates digital signatures as hash function factors, rendering third parties unable to compute the actual message sequence. Then for each segmentation, Alice computes one-time session key derived from a combination of stealth address and Bob's public keys through cryptographic hash function, which produces outputs that are evenly distributed across the entire range of possible values. Specifically, the private transaction key $r$ guarantees the randomness of session key. Subsequently, Alice uses one-time session key to encrypt message segments and perform XOR operation between ciphertexts and original stealth addresses, constructing new stealth addresses as outputs of *TransTx* with the generated transaction amount attached. We denote the new stealth addresses and masked amounts as a **special field of *TransTx***. The generation process of the special field for *TransTx* is described in Algorithm 1.

*3) State Feedback Method*

In order to mitigate the impact of decreased reliability of message transmission due to lost transactions, we design a state feedback method that allows receivers to inform senders about the receipt of message segments. If a blockchain attack prevents the sender's transaction, which contains the message, from reaching the receiver, the receiver will feedback on the missing message sequence. If all messages are received, the receiver will return feedback indicating a normal state to the sender. If the sender does not receive feedback after an extended period, he will detect the anomaly and implement remedial measures. In addition, the feedback provider embeds their digital signature in the transaction to ensure the traceability of the feedback.

This method is designed to enable receiver to provide feedback on the communication status. After Bob receives a *TransTx* with the transaction amount end flag indicating 0, he will assemble all extracted message segments and send feedback to Alice through a Monero transaction on whether any messages are missing. Bob embeds his digital signature into stealth address of Monero transactions. Additionally, as showing in Fig. 3, Bob inserts the missing message sequence number in the *FbTx* amount field. After receiving *FbTx*, Alice determines whether to resend the message based on the amount field of the *FbTx*. We denote the new stealth addresses and masked amounts as a **special field of *FbTx***. The generation process of the special field for *FbTx* is described in Algorithm 3.

IV. SECURITY ANALYSIS AND PERFORMANCE EVALUATION

This section presents a security analysis and evaluates our approach in terms of concealment, embedding rate and efficiency, and concludes with a comparative analysis against related works.

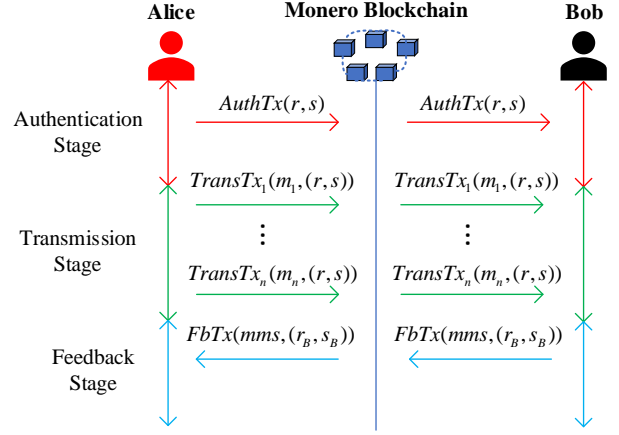

Fig. 2. Overview of Three Stages

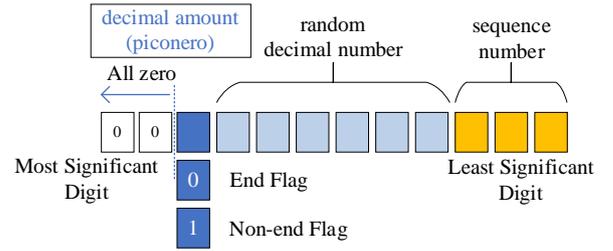

Fig. 3. Amount Encoding Method

---

**Algorithm 1** Generation Special Field of *AuthTx*

INPUT: Random factor $k^r$, Private view key of Alice $k_A^v$, Public view key of Bob $K_B^v$, Public spend key of Bob $K_B^s$

OUTPUT: Special field of *AuthTx* $sfa$

1. Compute $K^r = k^r G$
2. Compute $(r,s) = Sign(\mathcal{H}(K^r), k_A^v)$
3. Initalize special field of *AuthTx* $sfa = \{\}$
4. **for** $0 \leq i \leq 1$ **do**:
5.     Compute $K_i^{ori} = \mathcal{H}(k^r K_B^v, i)G + K_B^s$
6.     **if** $i == 0$ **then**
7.         Compute $K_i^{new} = r \oplus K_i^{ori}$
8.     **endif**
9.     **else**
10.        Compute $K_i^{new} = s \oplus K_i^{ori}$
11.     **endelse**
12.     $sfa$.append($K_i^{new}$)
13. **endfor**
14. Output $sfa$

**Algorithm 2** Generation Special fields of *TransTx*

INPUT: Message $m$, Public view key of Bob $K_B^v$, Public spend key of Bob $K_B^s$, Digital signature of Alice $(r,s)$

OUTPUT: Special fields of *TransTx* $sft$

1. Divide message $m$ into 32 bytes chuncks $m=\{m_1,m_2,...,m_n\}$
2. Initalize special field of AuthTx $sft=\{\}$
3. **for** $m_i$ in $m$ **do**:
4.    Get amount $a_i$ based on amount encoding method $a_i=GenAmount(i)$
5.    Choose a random factor $k_i^r$
6.    Compute public transaction key $K_i^r=k_i^r G$
7.    Initialize output index t as 0 or 1 as non-change out index
8.    Compute masked amount with digital signature of Alice $h_i=a_i\oplus_8 \mathcal{H}("amount",\mathcal{H}(k_i^r K_B^v,t),(r,s))$
9.    Compute original stealth address $K_i^{ori}=\mathcal{H}(k_i^r K_B^v,t)G+K_B^s$
10.   Compute one-time session key $k_i=\mathcal{H}(K_i^{ori}\|K_B^v\|K_B^s)$
11.   Compute $c_i=Enc(m_i,k_i)$
12.   Compute $K_i^{new}=K_i^{ori}\oplus c_i$
13.   Get transaction special field $sft_i=\{h_i,K_i^{new}\}$
14. **endfor**
15. Combine $sft=\{sft_1,sft_2,...,sft_n\}$
16. Output $sft$

**Algorithm 3** Generation Special fileds of *FbTx*

INPUT: Random factor $k^r$, Private view key of Alice $k_B^v$, Public view key of Bob $K_A^v$, Public spend key of Bob $K_A^s$, Missing message sequence $mms$

OUTPUT: Special field of *FbTx* $sff$

1. Compute $K^r=k^r G$
2. Compute $(r,s)=Sign(\mathcal{H}(K^R),k_B^v)$
3. Initialize special field of FbTx $sff=\{\}$
4. **for** $0\le i\le 1$ **do**:
5.    Compute $K_i^{ori}=\mathcal{H}(rK_B^v,i)G+K_A^s$
6.    **if** $i==0$ **then**
7.      Compute $K_i^{new}=r\oplus K_i^{ori}$
8.    **endif**
9.    **else**
10.      Compute $K_i^{new}=s\oplus K_i^{ori}$
11.   **endelse**
12.   $sff$.append($K_i^{new}$)
13. **endfor**
14. **if** $mms\ne 0$ **then**
15.   Get amount $a$ based on amount encoding method $a=GenAmount(mms)$
16.   Initialize output index t as 0 or 1 as non-change out index
17.   Compute hiding amount with digital signature of Bob $h_0=a\oplus_8 \mathcal{H}("amount",\mathcal{H}(k_0^r K_B^v,t),(r,s))$
18.   $sff$.append($h_0$)
19. **endif**
20. Get transaction sepecial field $sff=\{K_0^{new},K_1^{new},h_0\}$
21. Output $sff$

### A. Security Analysis

We now conduct a security analysis of the approach we proposed based on the **SRs** outlined in Section II.B.

#### 1) SR1: Confidentiality of on-chain session key

To ensure the confidentiality of session key, our proposed approach includes the following two considerations, which will be discussed separately.

Firstly, our approach does not require off-chain pre-negotiation of critical parameters of message transmission, such as master keys, encryption keys, etc. This allows the communicating parties to use publicly available wallet addresses for direct communication, reducing the impact of off-chain network attacks on covert transmissions.

Secondly, our approach can dynamically update session keys on-chain to encrypt messages, avoiding the risk of compromising associated sessions due to the leakage of long-term session keys. We demonstrate the confidentiality of the session keys and message through an assumption and specific scenarios.

*Assumption:* We assume there exists an attacker who knows all the public keys denoted in Table I and can associate a covert transaction containing messages with the public keys of both parties.

There are two scenarios in which the attacker can compute the dynamic key $k$, ultimately gaining access to the plaintext information.

*Scenario 1.* In this scenario, in order to get $K^{ori}$ by $K_B^v$, $K^r$ and $K_B^s$, based on the one-way and anti-collision property of hash functions, the attacker need to compute the shared secret $S$ shown in (5). Due to the difficulty of ECDLP, given $G$ and public keys, it is infeasible to compute $S$ without knowing the private keys, since the attacker would have to solve the ECDLP to find private keys.

*Scenario 2.* In this scenario, the attacker tries to extract original stealth address $K^{ori}$ and ciphertext $C$ from new stealth address $K^{new}$. Let $Adv_A^{xor}(s)$ denote the advantage of attacker $A$ in separating 256-bit $K^{ori}$ and ciphertext $C$ from 256-bit $K^{new}$, where the security parameter $s=256$. This advantage can be defined as the difference between the probability of the attacker successfully separating two strings and the probability of a random guess, shown in Eq. (8):

$$Adv_A^{xor}(s)=\Pr[A(K^{new})\rightarrow (K^{ori},C)]-\frac{1}{2^s}, \quad (8)$$

where $K^{new}=K^{ori}\oplus C$. Ideally, the attacker has no better chance than random guessing, i.e., $\Pr[A(K^{new})=(K^{ori},C)]$ is close to $\frac{1}{2^{256}}$, making $Adv_A^{xor}(s)$ close to zero, indicating that the advantage is negligible.

Based on the discussion above, our approach can avoid off-chain negotiations and ensures the confidentiality of dynamically updated session keys. Therefore, our approach satisfies **SR1**.

#### 2) SR2: Non-repudiation of transmission parties

Our approach ensures that while Monero transactions remain anonymous to third parties, the non-repudiation between the transmission parties still exists.

Firstly, during the authentication stage, the sender needs to embed their digital signature into the Monero transaction for the receiver, allowing the receiver to verify the validity of the signature. Secondly, during the transmission stage, the sender uses their signature in the process of generating masked amounts, ensuring that only the parties in possession of this signature can correctly extract the message sequence. Finally, our approach enables the receiver to embed their digital signature into the *FbTx* during the feedback stage, preventing the receiver from denying the communication behavior. Therefore, the communication parties' actions are accountable throughout the entire CT process, making it impossible to deny transmission behaviors. Thus, our approach meets **SR2**.

*3) SR3: Reliability of message transmission under blockchain attack*

The distributed nature of blockchain enhances the reliability of message transmission for blockchain-based CT. However, when the blockchain is under attack, such as eclipse attack, the transactions may be lost and not properly received by the intended receiver. Our approach offers mitigation measures for this scenario, addressing both the sender's and receiver's perspectives.

Firstly, the sender embeds message sequence and end flag in the masked amount when sending messages. This allows the receiver to assemble the complete message based the message sequence and end flag, thereby determining if the message is complete.

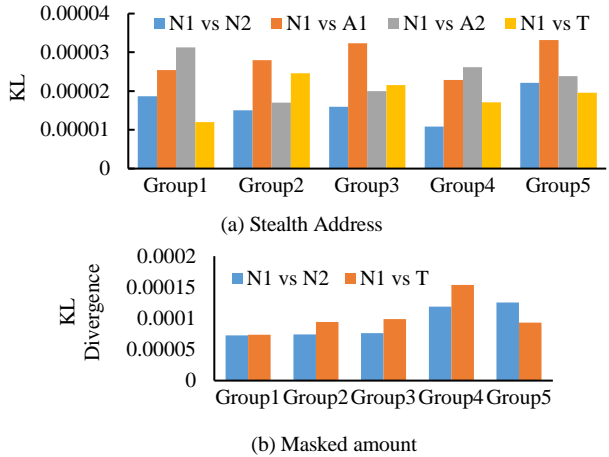

Fig. 4. KLD Experiments for Special Fields

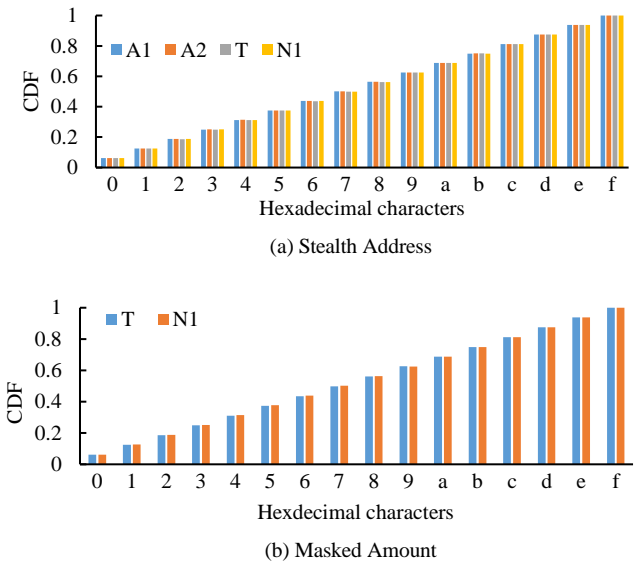

Fig. 5. CDF Experiments for Special Fields

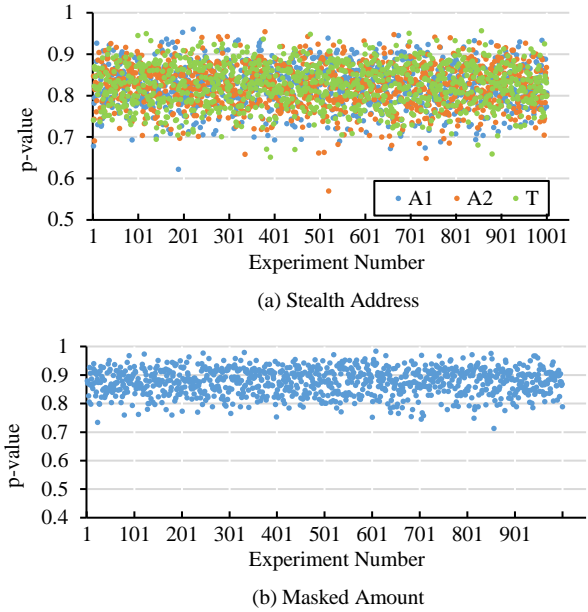

Fig. 6. KS Test for Special Fields

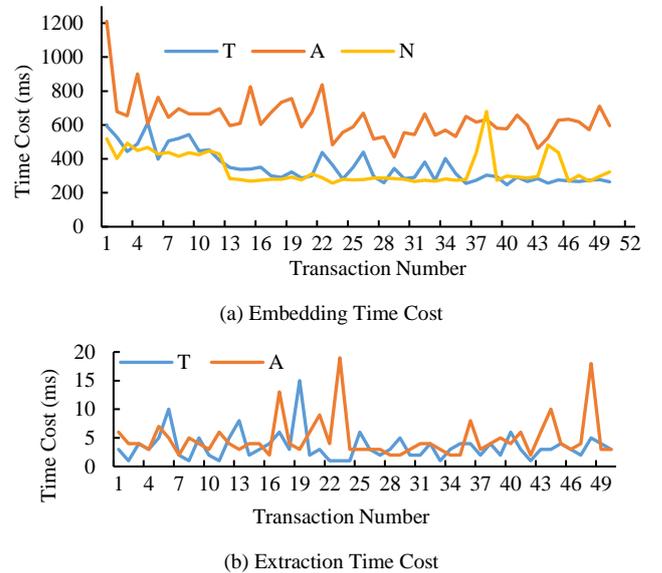

Fig. 7. Time Cost for Embedding and Extraction

TABLE V

AVERAGE TIME COST IN CREATION AND EXTRACTION PROCESS

|  | A | T | N |
|---|---|---|---|
| Average_creation (ms) | 639.10 | 352.56 | 339.74 |
| Average_extraction (ms) | 4.82 | 3.50 | - |

Secondly, upon receiving the whole message, the receiver assesses its completeness. after the assessment, the receiver sends a *FbTx* to the sender to inform the current communication status. Upon receiving *FbTx*, the sender decides whether to resend any message fragments. If the sender receives no *FbTx* after several new blocks, he will check their node's status and take remedial measures.

Our approach considers the reliability of message transmission from both the sender's and receiver's perspectives when under blockchain attacks, satisfying **SR3**.

TABLE IV
EXPERIMENTAL TRANSACTIONS ON MONERO STAGENET

| Transaction Type | ID | Block Height | Transaction Hash | Time Cost (ms) |
|---|---|---|---|---|
| Normal Transaction | 1 | 1589147 | a6e3910d40e53148464be53fb94a92a2550aaf98c4f3623a815c675964674de0 | 519 |
| | 2 | | 057ed40813ecfc3537000d44109e087369f501f0ebac7c33f625e2eeea85fc4a | 402 |
| | 3 | 1589148 | e5e1acf144f4baac6ba293c3c229bcd68582c80fe83fa036571ecfa4b51dd8f8 | 493 |
| | 4 | | 8f634d3f74122b2cff7187a4f2fe491881931af097e920fc632f83665ed4fe7e | 449 |
| | … | … | … | … |
| | 48 | | b74ce6d0271183b939508848aa254563054eb4ef7fcc2a37ea403b8e9cf87b0d | 268 |
| | 49 | 1589176 | a6ea59202b847e9cf2b23459c5d39e66a115258b4bafb64f25bab5856a7311c4 | 297 |
| | 50 | | e3955bdbc23e2530c25bb405e90d398b926cc706ac4e4aa8648574bf93b75be2 | 322 |
| Authentication Transaction | 1 | 1589042 | 2af16c992981f9ed7c9321e04bbce9250af1325176cb5ccb1e6f961adabf9c2b | 1211 |
| | 2 | | dc92de546d337f0d04afe5b0367ab2b3a656d55a5248a7c9827293ae7ab40308 | 678 |
| | 3 | 1589047 | 3362ac43a559437977f14be2e01e25a40a6ba48127ee938f3eaa1424874f6a4e | 655 |
| | 4 | | 3b9329b936dcde4825f0ef9e574790ce5bac526c030decedc0866dd269da60b9 | 901 |
| | … | | … | … |
| | 48 | | 124a2a9786eb00fdefa9d03f7f2a4ca922d6b4dd12d2756cbf58d6eba455e042 | 572 |
| | 49 | | d685a9b6d7ab10b477fd9f995287cb584f87b89b7801fee93da5f730aa375c12 | 710 |
| | 50 | 1589080 | 70b908cd02562035bd89ee6c7303fadd21d6a603c73256ccf9f60ddc0b2073b6 | 596 |
| Transmission Transaction | 1 | 1589094 | 764e7b8ec1cbe3f057fdece54c2d70bfafa46fd2ecb099d56e2a5a3cf90c8a31 | 594 |
| | 2 | 1589095 | 5ce93659549015badbb8b5ab3c666793f5b3ee98039f2e9fee3277aaf9398741 | 528 |
| | 3 | 1589096 | b376d9bd70f042223ed533bf8f5a1fdb69106e7c0670daeb55514c288ff7b0ae | 443 |
| | 4 | | bb09756d54b1206c70c3edd6221ec28469d59d2d34a25cb4cd209c526585c140 | 489 |
| | … | … | … | … |
| | 48 | | 74947780de112c551652ef5fd8812ce7eb8aaf751cd6788d83a53dd5d09b4288 | 274 |
| | 49 | 1589111 | 7cce5df0f9f1a0fc4d6ff28438d11736ad1231923453e76c635efdd315d11b65 | 278 |
| | 50 | | eb602d27f7211b07496eccb8f53dc57b3b9bbb173d816758868e4b5e8cb4e812 | 264 |

**4) SR4: Unlinkability of covert transactions**

Covert transactions should be unlinkable to mitigate the risk of statistical analysis due to increased transaction frequency. Unlinkability includes two aspects: one is that covert transactions cannot be directly linked to each other, and the other is that covert transactions should not be linkable through normal transactions.

In our approach, unlike Bitcoin and Ethereum, the ring signature mechanism and stealth address feature of Monero prevent third parties from obtaining the true public keys of the communicating parties from transaction data. Consequently, there is no direct or indirect link between covert transactions, fulfilling **SR4**.

**5) SR5: Obscurity of covert transactions**

Our approach allows the receiver to filter covert transactions using Monero stealth addresses without any obvious features, while ensuring that third parties cannot distinguish between special and normal transactions.

For the original stealth address and masked amount, they can be considered as a series of uniformly distributed hexadecimal strings. In our approach, we modify the stealth addresses and masked amount in Monero transactions to embed messages. Firstly, the new stealth address is calculated by $K^{new} = K^{ori} \oplus C$, where $C$ is the ciphertext encrypted with AES and can be also regarded as randomly distributed. This does not alter the randomness of stealth address. Secondly, the modified masked amount is generated by cryptographic hash function and is uniformly distributed as well. Therefore, special fields and normal fields are statistically indistinguishable, ensuring the stealthiness of communication and thus our approach satisfies **SR5**. The experiments of indistinguishability between special and normal transactions are provided in Section IV.0.

*B. Performance Evaluation*

We implemented our approach on Monero==0.18.1.0, running in an Ubuntu 22.04 virtual machine hosted on a PC with Windows 11 operating system, equipped with an Intel i9-13900HX processor and 32 GB memory.

To conduct our experiments, we wrote configuration files for both the Monero node and wallet, deploying them to the Ubuntu 22.04 virtual machine, connecting the Monero full node to the Monero stagenet for experimental purposes. Thanks to the low mining difficulty on stagenet, we were able to use CPU mining to obtain spendable Monero coins.

Additionally, we developed a message handler to automatically handle messages and encode transaction amounts, which is used to automate requests to the Monero wallet RPC interface and collect experimental data.

*1) Concealment*

In our approach, we transmit digital signature and message segments by generating special stealth address of Monero transactions. In addition, transaction amount fields are used to cover real amount, which are generated according to message sequences or missing message orders. The concealment of covert transactions is defined by their indistinguishability between ordinary and special fields. To comprehensively assess the indistinguishability, we employ three statistical methods: Kullback-Leibler Divergence (KLD), Cumulative Distribution Function (CDF), and Kolmogorov-Smirnov Test (KS Test) [13][15][22][25]. In these three types of experiments, we constructed a large number of transactions to evaluate concealment. Since the evaluation only requires data of special and normal fields, these transactions were not broadcast to the Monero network. This did not affect the correctness of experiment results.

KLD is a statistical measure used to quantify the difference between two probability distributions. When two random distributions are the same, their KLD is zero. As the difference between the two distributions increases, their relative entropy also increases. Eq. (9) gives the calculation of KLD, where $P(x)$ and $Q(x)$ represent the probability of event $x$ in the two probability distributions, respectively.

$$D_{KL}(P \| Q) = \sum_x P(x) \log \frac{P(x)}{Q(x)} \qquad (9)$$

We first use KLD to evaluate the normal fields and special fields embedded with message and signature. We conduct KLD experiments between normal and special fields of *AuthTx*, *TransTx* and normal transactions, including stealth addresses and masked amounts. Due to the similarities in the processing procedures of *FbTx* and *AuthTx*, the specific experiments for *FbTx* are omitted. In detail, we generated 50,000 *AuthTx*, *TransTx*, and normal transactions initially, each divided into five groups. For *AuthTx*, we recorded the stealth addresses of two non-change outputs, named *A1* and *A2*. For *TransTx* and normal transactions, we saved the stealth addresses and masked amounts of one non-change output, named *T* and *N1*. To be clearer, we additionally generated 10,000 standard stealth addresses and masked amounts named *N2*. This helps identify any anomalies in statistical analysis and verify the significance of the results. Then, we calculate the frequency distribution of hexadecimal characters for A1, A2 and N1, and compare it with the character frequency distribution of N2 to compute the KLD.

Fig. 4 shows the result of KLD of stealth addresses and masked amounts, which reveals that the KLD is uniformly at the order 5e-10 and 4e-10 separately, and there is essentially no difference in KLD between the five groups. Small and similar KLD across groups indicate that the data distributions are nearly identical. Therefore, the stealth addresses between *AuthTx*, *TransTx*, and normal transactions and the masked amounts between *TransTx* and normal transactions are indistinguishable.

Subsequently, we conduct CDF experiment to compare character frequency between standard and special fields. CDF represents the probability that a random variable takes on a value less than or equal to a specific number. As what we have done in KLD, we collect character frequency in stealth addresses and masked amounts of *AuthTx*, *TransTx* and normal transactions. For stealth addresses, we calculate CDFs of *A1*, *A2*, *T* and *N1*. For masked amounts, we calculate CDFs of T and N1. The result is showed in Fig. 5, in which the CDFs for all groups appear to increase linearly, suggesting that the frequency distribution of hexadecimal characters is uniform across the normal and special fields for each group. The CDF of each group is nearly identical at the same character position, demonstrating the consistency of character distribution.

KS Test is designed to compare two samples to determine if they are drawn from the same distribution. We perform a KS Test on four sets of data previously depicted, including those with stealth addresses and masked amounts to evaluate indistinguishability. For each set, we conduct 1000 samples, with each sample consisting of 500 instances. We calculated p-values for the character frequency comparison between normal and special fields, and the result are displayed in Fig. 6. It shows that almost all p-values appear to fall between 0.6 and 0.8, which is higher than the significance level of 0.05, suggesting that for most of the samples, there is no statistical difference between the normal and special fields.

Based on the three types of experiments described above, the results indicate that the modified stealth address and masked amount fields in our approach cannot be distinguished from the original fields, demonstrating the concealment of our approach.

*2) Embedding Rate*

The embedding rate refers to the amount of information that can be carried in one covert transaction. Each 256-bit message is encrypted with a dynamic session key to generate a 256-bit ciphertext, which is then embedded into a 256-bit stealth address. Therefore, each *TransTx* is capable of transmitting a 256-bit message.

*3) Efficiency*

To evaluate the efficiency of our approach, we constructed and broadcast covert transactions, and extracted digital signatures and messages through querying transaction hashes. After sending an embedding request, the Monero client constructs and broadcasts transactions with embedded digital signature or messages, and returns a response upon successful broadcast. After sending an extraction request, the Monero client queries the transaction details using the transaction hash, extracts the digital signature of messages, and returns a response upon successful extraction. We define the time consumption as the difference between the response timestamp and request timestamp.

Firstly, we created 50 instances each of normal transactions, *AuthTx*, and *TransTx*, recording the corresponding times $T_{cost}$ as N, A, and T respectively, as shown in Fig. 7. (a). Secondly, we extracted digital signature and message from the 50 *AuthTx* and *TransTx* instances, respectively. The time consumed for these processes is also denoted as A and T, as illustrated in Fig. 7. (b). We list some of these broadcasted transaction hashes and corresponding blocks in TABLE IV, which can be found in Monero stagenet explorer online. The average time cost is presented in TABLE V. Combining Fig. 7 and TABLE V, it is evident that *AuthTx* require the most time due to the need for two non-change outputs, while *TransTx* and normal transactions involve one non-change output each. Furthermore, the construction and broadcast time for *TransTx* is about 12.82ms longer compared to a normal transaction, which is within an acceptable range. For the extraction process, the time required to extract signatures and information is very short, demonstrating high extraction efficiency.

*4) Comparison*

Our approach aims to provide a high embedding rate CT channel with no observable characteristics, which could dynamically update session key without any off-chain negotiation.

TABLE VI shows a comparison of various CT methods. BLOCCE embeds covert information into the last digit of Bitcoin address. Although it has provable security, its practical value is limited due to a low embedding rate. MRCC can embed 11 bits per input in a Monero transaction, and this capacity increases with the number of inputs. But in the Monero network, transactions with a large number of inputs are uncommon. Additionally, there exists a risk of sender denial of communication behavior in this method. Both DLchain, RDSAC, as well as our method, have an embedding rate of 256 bits per transaction, but the first two require off-chain negotiation of encryption keys, while our approach dose not. Moreover, due to the anonymity features of Monero, the public keys of both parties do not appear in plaintext within the Monero network, which grants our system a higher level of concealment compared to other systems.

TABLE VI

COMPARISON OF DIFFERENT CT APPROACH

| Approach | Embedding Rate (bit/tx) | Dynamic Session Key | Observable characteristic | Off-chain Negotiation |
|---|---|---|---|---|
| BLOCCE [6] | 1 | No | Yes | Yes |
| DLchain [7] | 256 | No | Yes | Yes |
| MRCC [18] | $11r$ | No | No | Yes |
| RDSAC [25] | 256 | No | No | Yes |
| Ours | 256 | Yes | No | No |

## V. CONCLUSION

In this paper, we propose a novel CT approach named as MBCT, which updates session key on-chain dynamically without off-chain negotiation. We first present a threat model and security requirements for blockchain-based CT. Then, MBCT is detailed and we make security analysis to validate that MBCT can meet the requirements. Finally, we implement MBCT in Monero-0.18.1.0 and the experiment results demonstrate its high embedding capacity.

REFERENCES

[1] K. Lu, H. Liu, L. Zeng, J. Wang, Z. Zhang and J. An, "Applications and prospects of artificial intelligence in covert satellite communication: a review," *Science China Information Sciences*, 2023, 66(2): 121301.

[2] X. Chen et al., "Covert Communications: A Comprehensive Survey," *IEEE Communications Surveys & Tutorials*, vol. 25, no. 2, pp. 1173-1198, Secondquarter 2023.

[3] J. An, B. Kang, Q. Ouyang, J. Pan and N. Ye, "Covert Communications Meet 6G NTN: A Comprehensive Enabler for Safety-Critical IoT," *IEEE Network*, early access, doi: 10.1109/MNET.2024.3379864.

[4] P. Yang, Y. Li and Y. Zang, "Detecting DNS covert channels using stacking model," *China Communications*, 2020, 17(10): 183-194.

[5] Z. Chen et al., "Blockchain Meets Covert Communication: A Survey," *IEEE Communications Surveys & Tutorials*, vol. 24, no. 4, pp. 2163-2192, Fourthquarter 2022.

[6] Partala J, "Provably secure covert communication on blockchain," *Cryptography*, 2018, 2(3): 18.

[7] J. Tian, G. Gou, C. Liu, Y. Chen, G. Xiong and Z. Li, "DLchain: A covert channel over blockchain based on dynamic labels," in *Proc. of International Conference on Information and Communications Security*, pp. 814-830, 2019.

[8] W. Wang and C. Su, "CCBRSN: A system with high embedding capacity for covert communication in Bitcoin," in *Proc. of International Conference on ICT Systems Security and Privacy Protection*, pp. 324-337, 2020.

[9] Z. Wang et al., "A covert channel over blockchain based on label tree without long waiting times," *Computer Networks*, vol. 232, Aug. 2023.

[10] X. Luo, P. Zhang, M. Zhang, H. Li and Q. Cheng, "A Novel Covert Communication Method Based on Bitcoin Transaction," *IEEE Transactions on Industrial Informatics*, vol. 18, no. 4, pp. 2830-2839, April 2022.

[11] X. Zhang, X. Zhang, X. Zhang, W. Sun, R. Meng and X. Sun, "A derivative matrix-based covert communication method in blockchain," *Computer Systems Science and Engineering*, vol. 146, no. 1, pp. 225-239, 2023.

[12] H. Cao et al., "Chain-Based Covert Data Embedding Schemes in Blockchain," *IEEE Internet of Things Journal*, vol. 9, no. 16, pp. 14699-14707, 15 Aug.15, 2022.

[13] L. Zhu, Q. Liu, Z. Chen, C. Zhang, F. Gao and Z. Yang, "A Novel Covert Timing Channel Based on Bitcoin Messages," *IEEE Transactions on Computers*, vol. 72, no. 10, pp. 2913-2924, Oct. 2023.

[14] A. I. Basuki and D. Rosiyadi, "Joint transaction-image steganography for high capacity covert communication," in *Proc. Int. Conf. Comput. Control Informat. Appl.*, pp. 41-46, 2019.

[15] S. Liu et al., "Whispers on Ethereum: Blockchain-based covert data embedding schemes," in *Proc. 2nd ACM Int. Symp. Blockchain Secure Crit. Infrastruct.*, pp. 171-179, Oct. 2020.

[16] L. Zhang, Z. Zhang, Z. Jin, Y. Su and Z. Wang, "An approach of covert communication based on the Ethereum whisper protocol in blockchain," *International Journal of Intelligent Systems.*, vol. 36, no. 2, pp. 962-996, Feb. 2021.

[17] L. Zhang, Z. Zhang, W. Wang, Z. Jin, Y. Su and H. Chen, "Research on a Covert Communication Model Realized by Using Smart Contracts in Blockchain Environment," *IEEE Systems Journal*, vol. 16, no. 2, pp. 2822-2833, June 2022.

[18] Z. Guo, L. Shi, M. Xu and H. Yin, "MRCC: A Practical Covert Channel Over Monero With Provable Security," *IEEE Access*, vol. 9, pp. 31816-31825, 2021.

[19] L. Liu, L. Liu, B. Li, Y. Zhong, S. Liao and L. Zhang, "MSCCS: A Monero-based security-enhanced covert communication system," *Computer Networks.*, vol. 205, Mar. 2022.

[20] W. Su and L. Ma, "A Blockchain-based Covert Document Communication System Model," in *2023 8th International Conference on Computer and Communication Systems*, Guangzhou, China, 2023, pp. 445-450.

[21] F. Gao, L. Zhu, K. Gai, C. Zhang and S. Liu, "Achieving a covert channel over an open blockchain network," *IEEE Network.*, vol. 34, no. 2, pp. 6-13, Mar. 2020.

[22] C. Zhang, L. Zhu, C. Xu and R. Lu, "EBDL: Effective blockchain-based covert storage channel with dynamic labels," *Journal of Network and Computer Applications*, 2023, 210: 103541.

[23] J. Liu et al., "DLCCB: A Dynamic Labeling Based Covert Communication Method on Blockchain," in *2023 International Wireless Communications and Mobile Computing*, Marrakesh, Morocco, 2023, pp. 168-173.

[24] P. Zhang, Q. Cheng, M. Zhang and X. Luo, "A Group Covert Communication Method of Digital Currency Based on Blockchain Technology," *IEEE Transactions on Network Science and Engineering*, vol. 9, no. 6, pp. 4266-4276, 1 Nov.-Dec. 2022.

[25] Z. Chen, L. Zhu, P. Jiang, C. Zhang, F. Gao and F. Guo, "Exploring Unobservable Blockchain-based Covert Channel for Censorship-Resistant Systems," *IEEE Transactions on Information Forensics and Security*, 2024.

[26] P. Zhang, Q. Cheng, M. Zhang and X. Luo, "A Blockchain-Based Secure Covert Communication Method via Shamir Threshold and STC Mapping," *IEEE Transactions on Dependable and Secure Computing*, early access, doi: 10.1109/TDSC.2024.3353570.

[27] Alonso K M, "Zero to monero," *Zero to monero*, 2020.

[28] G. Kappos, H. Yousaf, M. Maller and S. Meiklejohn, "An empirical analysis of anonymity in Zcash," in *Proc. 27th USENIX Conf. Security Symp.*, pp. 463-477, 2018.

[29] F. Liu et al., "Bitcoin Address Clustering Based on Change Address Improvement," *IEEE Transactions on Computational Social Systems*, early access, doi: 10.1109/TCSS.2023.3239031.

[30] A. Mansourabady, F. Tabe, A. H. Rasekh and A. Ghermezian, "A Study on Hybrid Deep Learning Approaches for "Monero" Cryptocurrency Price Prediction," in *2024 20th CSI International Symposium on Artificial Intelligence and Signal Processing (AISP)*, Babol, Iran, Islamic Republic of, 2024, pp. 1-6.